# Non-linear optical study of hierarchical 3D Al doped ZnO nanosheet arrays deposited by successive ionic adsorption and reaction metod

Ilyass Jellal a, Khalid Nouneh a,∗, Jaroslaw Jedryka b,∗, Denis Chaumont c, Jamal Naja d

a Laboratory of Physics of Condensed Matter (LPMC), Department of Physics, Faculty of Sciences, Ibn Tofail University, BP 133 Kenitra, Morocco

b Institute of Optoelectronics and Measuring Systems, Faculty of Electrical Engineering, Czestochowa University of Technology, 17 ArmiiKrajowej Str., 42-200, Czestochowa, Poland

c NanoForm, Laboratoire Interdisciplinaire Carnot de Bourgogne ICB (UMR 6303 CNRS), Université de Bourgogne Franche-Comté, Dijon, France

d Applied Chemistry and Environment, Faculty of Science and Technology, Hassan 1st University, B.P 577 Route de Casablanca, Settat, Morocco

## Abstract

Successive ionic layer adsorption and reaction (SILAR) method is based on the adsorption and reaction of the ions in the cationic solution and the ionic solution, respectively. This method is simple, inexpensive, large-scale deposition, effective way for deposition on 3D substrates, low-temperature process and represents an easy way for the preparation of doped, composite and heterojunction materials. To take advantage of this method and the ZnO nanostructures, various parameters have been optimized. Undoped and Aluminum (Al) doped ZnO nanostructures were prepared by the SILAR technique. The characterization of the nanostructures prepared was carried out using X-ray diffraction (XRD), scanning elektron microscopy (SEM), energy dispersive spectroscopy (EDS), X-ray photoemission spectroscopy (XPS), UV–Vis spectrometry and nonlinear optical analysis (NLO). The structural, compositional and optical properties confirm the introduction of Al3+ ions into the ZnO matrix. As a result, an enhancement of the crystallinity, enhancement of the light absorption and a change in the morphology of the nanostructures were observed. The laser stimulated nonlinear optical effects of the second and third harmonic generation were done using a fundamental laser beam. The laser stimulated NLO values obtained are at least 10% higher than the doped ZnO nanomaterials synthesized by other methods using the same set-up.

## 1. Introduction

Recently, an enhanced interest is observed in the study of wellknown ZnO nanostructures for application in different optoelectronic devices [1–4]. Among a lot of optoelectronic materials, zinc oxide Has significant importance due to its interesting physical and chemical properties: a large direct energy band gap (about 3.37 eV), a high excitonic binding energy (60 meV). It can be applied in light-emitting diode (LEDs), solar cells [5], piezoelectric devices, modulators and sensors [6], biosensors [7], immunosensors [8] and photocatalytic degradation [9]. Furthermore, ZnO nanostructures may be promising for applications in the field of nonlinear optics such as optical limiters, saturable absorbers used for passive Q-switching, and optical modulators and deflectors [10–13]. Optical nonlinearity appears in the medium through which the laser light is propagated and it is typically observed only at higher power densities generated by lasers. The nonlinearity crucially depends on the natural medium. The wurtzite-like ZnO crystals are noncentrosymmetry and have a good NLO efficiency described by third order polar tensor. Additionally, they have a higher laser damage threshold with respect to the other semiconductors making it an appropriate material for NLO applications both in the bulk and nanocrystalline form [14]. ZnO nanostructures exhibit second harmonic generation (SHG) and also third order nonlinear optical effects, such as two photon absorption and third harmonic generation (THG). These NLO features are sensitive to the nanofilms orientation as well as intrinsic and extrinsic defects inside the crystal structure. Hence to tailor the NLO properties, different doping elements are introduced to the crystal lattice of ZnO nanostructures [11,13,15,16]. The particular impurity energy levels form the localized energy levels within the energy gap permitting additional manipulation by their features Rusing external fields. For instance, an enhancement of the THG upon an increase in irradiation by electron dose rates was observed [17]. Additionally, these materials may be applied also as photocatalyst and sensors due to their promising photocatalytic, piezoelectric, optical features [9,18–20]. These properties of the ZnO nanostructures will be crucially dependent on the technique used. The zinc oxide crystalline nanostructures can be fabricated using different methods, including sol–gel, pulsed laser deposition, sputtering, spray pyrolysis, hydrothermal and SILAR. SILAR processing is a relatively new and less studied method [21]. Its advantages are based on simplicity, low cost of operation and adaptation to the deposition of uniform thin films on large size substrates [22]. Moreover, it can be used for deposition of well-covering films on 3D substrates, because the liquid is in contact with the en tire active surface of the substrate. This technique is based on the adsorption and reaction of the ions from the solutions (adsorption of cations followed by reaction with anions) [23]. The rinsing step is used between every immersion to avoid precipitation in the solution and also to eliminate unwanted species [24]. A cycle of this technique is defined when a substrate passes by this four-step. The schematic set-up of the SILAR method is presented in Fig. 1. In the case of the modified SILAR method, the rinsing step was avoided to increase the rate of deposition and to reduce deposition time [25]. In the present work, nanostructures of zinc oxide were prepared using the modified SILAR method. The doping of ZnO, may be necessary to improve its optical and electrical properties [26]. Aluminum is the best candidate due to its non-toxic nature, low-cost and good electrical conductivity. Al moped ZnO has a high electrical conductivity due to the substitution of $Zn^{2+}$ by $Al^{3+}$ which may be promising for the NLO [19,27]. The introduction of Al into ZnO has been reported to affect the third harmonic generation [28]. Maximum THG values were obtained for 1% Al and suppression of THG values were obtained for higher Al content (3% and 5%). Al moped ZnO nanostructures have many advantages such as the large surface/ volume ratio and present a promising candidate for the fabrication of optical devices at the nanometric scale [29]. The nature of the precursors influences the growth of the material [30]. It has been reported [31] that the SILAR technique is used to prepare Al doped ZnO using chloride-based precursors and the morphology obtained is nanopebbles, which do not uniformly cover the substrate.

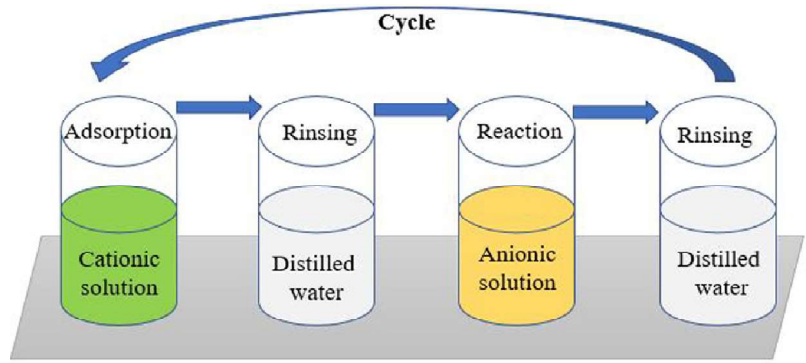

Fig. 1. Schematic representation of the SILAR metod

When sulfate-based precursors were used for preparing Al. doped ZnO by SILAR, large crystallites were obtained [32]. Due to the high interest to the Al doped ZnO nanostructures and to benefit the advantages of the SILAR method, we have optimized all the parameters of this technique to obtain Al doped ZnO nanostructures for nonlinear optical studies.

2. Material and methods

The modified SILAR approach is used to prepare nanostructures of undoped and Al doped ZnO on a glass substrate. Zinc Sulfate Heptahydrate (ZnSO4,7H2O) was dissolved in ultrapure water (100 ml) to obtain a 0.08 M solution. Complexing agent (NH4OH) was added for adjusting the pH and forming the zinc ammonia complex, which can be easily adsorbed on the substrate. The immersion time of each step was 30 s. Firstly, a clean substrate was immersed in a zinc ammonia complex solution. Secondly, the substrate was transferred to another beaker containing ultrapure water maintained at 80 °C. Wherein the zinc ammonia complex was transformed into zinc hydroxide (Zn(OH)2) according to the Eqs. (1) and (2).

$$[Zn(NH_3)_4]^{2+} + 4H_2O \rightarrow Zn^{2+} + 4NH_4^+ + 4OH^- \qquad (1)$$

$$Zn^{2+} + 2OH^- \rightarrow Zn(OH)_2 \qquad (2)$$

According to Eq. (3), the zinc hydroxide can be transformed into a zinc oxide phase at ambient temperature.

$$Zn(OH)_{2(s)} \rightarrow ZnO + H_2O \qquad (3)$$

To complete the ZnO formation, samples are heat treated (500 °C) in a furnace under air atmosphere (Eq. (4)).

$$Zn(OH_2)_{(s)} + O_{2(g)} \xrightarrow{T(500°C)} ZnO_{(g)} + H_2O + O_{2(g)} \qquad (4)$$

Finally, an extremely adherent film of zinc oxide was obtained. To study the effect of aluminum doping, aluminum sulfate (Al2(SO4)3) was added to the previous mixture with different Al/Zn molar ratios: 2% and 4%. All the films were deposited as follows: 30 cycles done followed with heat treatment at 500 °C for 2 h. The crystal structure of the nanostructures was determined by X-Ray Diffraction (XRD) (BRUKER-D2 PHASER), with CuKα radiation of 1.54056 Å over the range within 20–80° with a recording step of 0.02° and a scanning step time of 165 s. The instrumental error of the XRD diffractogram is ± 0.02° of

2θ over the entire angular range. This value is used to calculate the error on crystallite sizes determined by Scherrer's formula and it was found to be around 2 nm as mentioned earlier [33]. The error in calculating lattice parameters using Bragg's law was determined and it was found to be around 0.004 Å and 0.006 Å for a and c parameters, respectively. The UV–Visible properties were recorded using a Spectronic Helios Gamma UV–VIS spectrophotometer over the range within 300–1000 nm with 0.5 nm step. It should be noted that the band gap of direct transition material determined Rusing Tauc's plot can be quite accurate, to ~1% [34]. This indicates that the error in the ZnO band gap value determined by Tauc's plot is around ± 0.033 eV [35]. The surface morphology of the samples was monitored by Scanning Electron Microscopy (SEM) (Hitachi SU8230). The surface chemical compositions were studied by Energy Dispersive Spectroscopy (EDS) using Oxford detector for an analysis area of 10 × 10 μm2. The EDS detector error is approximately ± 2 at.% [36]. X-ray Photoemission Spectroscopy (XPS) spectra were recorded by a PHI Versaprobe 5000 instrument. XPS analysis was carried out using Al. Kα1 X-rays (energy = 1486.6 eV, recording step = 0.05 eV, power = 50 W and diameter = 200 μm). XPS survey spectra on the surface without etching was produced. C 1s peak with binding energy (B.E) of about 284.8 eV was used for calibration. Fig. 1. Schematic representation of the SILAR method. For nonlinear optical studies, a 10 ns Nd:YAG fundamental pulsem laser generating at 1064 nm (pulse time equal to about 8–10 ns) with frequency repetition 10 Hz was applied as a fundamental light for the laser stimulated nonlinear optical studies. The principal set-up used in this study is shown in Fig. 2.

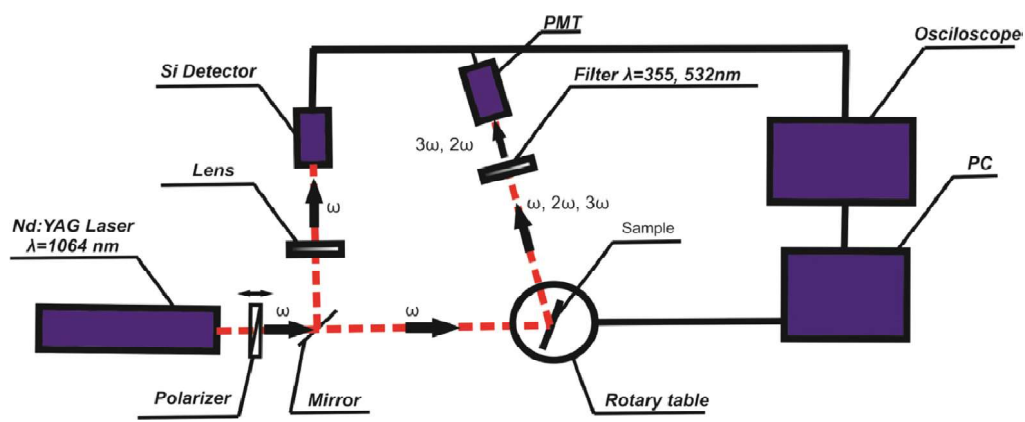

Fig. 2. Principal NLO set-up for monitoring of the laser induced SHG and THG

To tune the output fundamental energy density, a Glan polarizer was used. Semi-transparent mirrors together with a couple of lenses have been applied for the space split of the fundamental beam for Two coherent channels. One of the channels has been served (at the same frequency) for the monitoring of the fundamental laser beam control and another one for the probing of the SHG and THG after their spectra selection by interference filters at 532 nm and 355 nm, respectively. Furthermore, the spectral contours of the SHG/THG for the double and tripled frequency beams have been detected by spectrometer DFS8 (with diffraction gratings) with spectral resolution about 0.2 nm. This was done to eliminate the fluorescence parasitic background. The incydent fundametnal laser beam allowed to perform the laser coherent treatment during several seconds to form some anisotropic grating similarly to the described in [37].

### 3. Results and discussion

## 3.1. Structural, morphological and compositional properties

The X-ray diffraction spectra of undoped and Al doped ZnO structures are shown in Fig. 3. The films are polycrystalline and correspond to the wurtzite structure (DATA N°. 01-089-7102). An increase in the intensities of the peaks was observed when the doping process was applied. The growth along the c-axis was improved in the presence of aluminum. The increase in the intensity of the peaks indicates an improvement in the crystallinity of the films without affecting the Basic hexagonal wurtzite structure. The crystallite sizes (D) of the films was evaluated by the use of Scherrer formula (Eq. (5)) [38].

$$D = \frac{0.9\lambda}{\beta cos\theta} \quad (5)$$

$\lambda$, $\beta$ and $\theta$ are the X-ray wavelength ($\lambda$ = 1.54059 Å), the broadening of the diffraction line measured at half of its maximum intensity in radians and the angle of diffraction, respectively. The values calculated are presented in Table 1. The lattice parameters (a and c) were calculated using Bragg's law for the hexagonal system using Eqs. (6) and (7) [39].

$$a = \frac{\lambda}{\sqrt{3} \cdot sin\theta} \quad (6)$$

$$c = \frac{\lambda}{sin\theta} \quad (7)$$

Table 1 presents the structural parameters calculated for undoped and Al doped ZnO nanostructures. The variation of parameters D and a remained within the error limit of the formulas used to calculate these parameters.

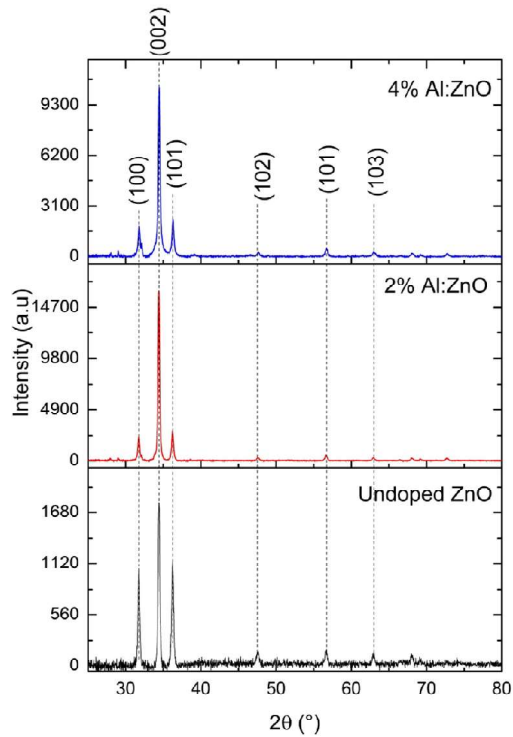

Fig. 3. XRD spectra of undoped and Al doped ZnO nanostructures

However, the variation in parameter c was significant. In the presence of aluminum, the c values slightly decreased. This is probably due to the substitution of Al3+ to Zn2+ sites, which is due to the difference in the ionic size of Al3+ (54 pm) and Zn2+ (74 pm) [40]. The SEM micrographs of undoped and Al doped ZnO nanostructures are shown in Fig. 4. The undoped ZnO sample has a morphology of nanocrystallites with flower-like morphological structures as displayed in Fig. 4a. These structures are rod-like with an average diameter varying within 150–200 nm and their length is equal to approximately 1 μm. As can be seen from Fig. 4b and 4c, Al doping resulted in a morphological change from rod-like structures to hierarchical 3D nanosheets structures. The cross section (Fig. 4d and 4e) shows that the thickness increased from 200 nm to 8 μm for undoped and 2% Al:ZnO, respectively.

Tab. 1. Structural properties of undoped and Al doped ZnO nanostructures

| Sample | Pic | D(± 2 nm) | Average D (± 2 nm) | a (± 0.004 Å) | c (± 0.006 Å) | c/a |
|---|---|---|---|---|---|---|
| Undoped ZnO | (1 0 0) | 31 | 30 | 3.25 | 5.211 | 1.603 |
|  | (0 0 2) | 32 |  |  |  |  |
|  | (1 0 1) | 27 |  |  |  |  |
| 2% Al:ZnO | (1 0 0) | 30 | 31 | 3.251 | 5.206 | 1.601 |
|  | (0 0 2) | 33 |  |  |  |  |
|  | (1 0 1) | 30 |  |  |  |  |
| 4% Al:ZnO | (1 0 0) | 30 | 31 | 3.247 | 5.202 | 1.602 |
|  | (0 0 2) | 32 |  |  |  |  |
|  | (1 0 1) | 31 |  |  |  |  |
| DATA 01-089-7102 | – | – | – | 3.249 | 5.206 | 1.602 |

This result confirms that the crystallization of the ZnO material is stimulated in the presence of aluminum. Table 2 presents the EDS analysis of undoped and Al doped ZnO nanostructures. EDS results of undoped ZnO confirmed the presence of Zn and O with an atomic percentage close to the stoichiometric ratio (47 at.% of Zn and 53 at.% of O). In the case of Al doped ZnO nanostructures, the presence of Zn, O and Al atoms was confirmed. The increase of the Al molar amount in the solution leads to an increase in the aluminum atoms in the films. The variation of O at.% remained within the error limit of EDS analysis, which indicates that the atomic % of O remained almost constant for all the samples, while the atomic % of Zn decreased when the atomic % of Al increased.

Tab. 2. EDS results of undoped and Al doped ZnO nanostructures

| Sample | O at.% | Zn at.% | Al at.% |
|---|---|---|---|
| Undoped ZnO | 53 | 47 | 0 |
| 2% Al:ZnO | 52 | 45 | 3 |
| 4% Al:ZnO | 53 | 42 | 5 |

XPS survey spectra on the surface for undoped ZnO and 2% Al:ZnO was produced (Fig. 5). The core line of Zn 2p3/2 of undoped ZnO and 2% Al:ZnO is shown in Fig. 5a.

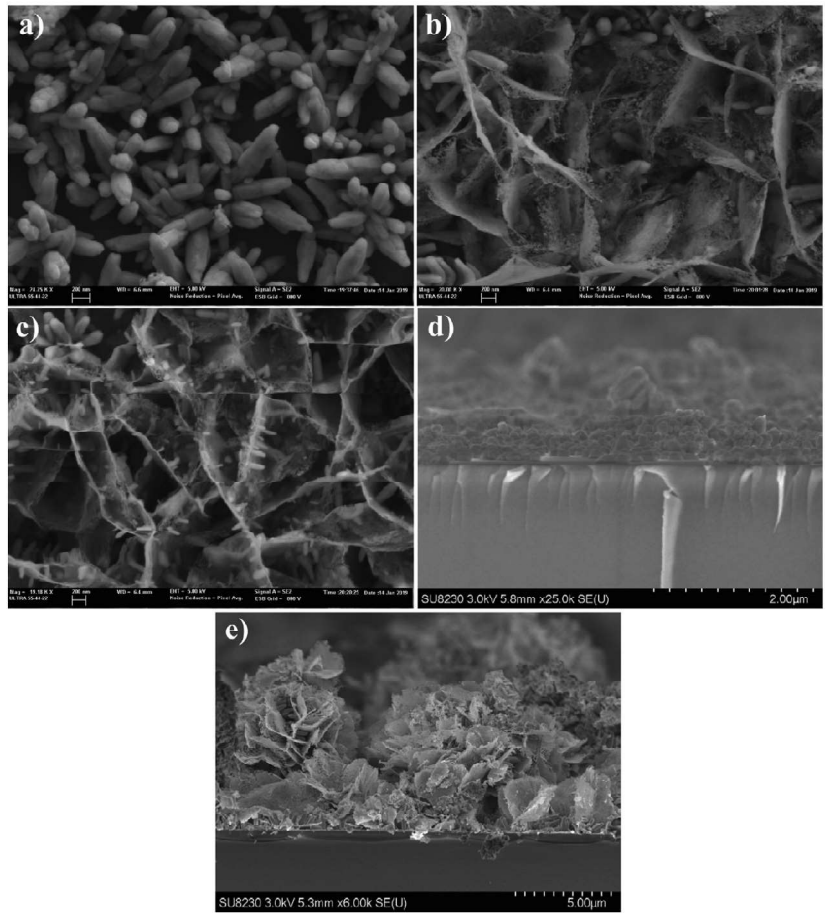

Fig. 4. SEM images of undoped and Al doped ZnO nanostructures: (a) undoped ZnO, (b) 2% Al:ZnO and (c) 4% Al:ZnO, and cross section of: (d) undoped ZnO, and (e) 2% Al:ZnO

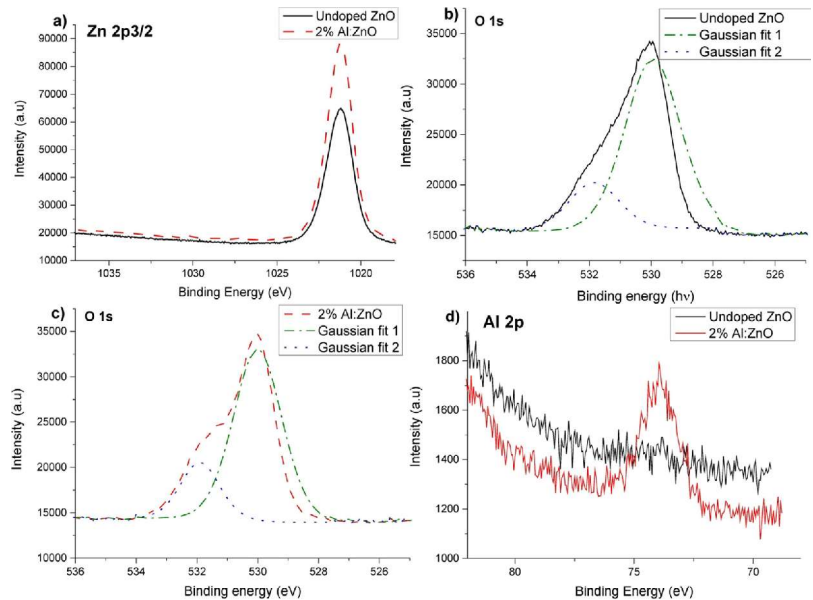

Fig. 5. XPS spectra of undoped ZnO and 2% Al:ZnO nanostructures, (a) core line of Zn 2p3/2, (b) Gaussian fit of O 1s peak of undoped ZnO, (c) Gaussian fit of O 1s peak of 2% Al:ZnO and (d) core line of Al 2p

The Zn 2p3/2 peak intensity increased when Al doping was applied. Fig. 5b and 5c present the O 1s peak of undoped ZnO and 2% Al:ZnO nanostructures, respectively. These spectra had almost the same intensity and were concentrated at a binding energy of about 530 eV. The peak O 1s of undoped ZnO and 2% Al:ZnO presented an asymmetrical feature which indicates the possibile presence of multi-component of O. Gaussian peak fitting was applied to search oxygen nature. The O 1s peak of undoped ZnO and 2% Al:ZnO can be resolved into two components by Gaussian fit as it's shown in Fig. 5b and Fig. 5c, respectively. The two peaks obtained were located at around 530 eV (Gaussian fit 1) and 531.9 eV (Gaussian fit 2). The Al. 2p peak of undoped ZnO and 2% Al:ZnO is illustrated in Fig. 5d. In the case of undoped ZnO, no peak of Al 2p was observed. On the contrary, one peak observed at a B.E of 74 eV for 2% Al:ZnO. XPS spectra confirmed the presence of Al, Zn, C and O in the samples analyzed on the surface. No peak was observed at the Winding energy 1021.5 eV which excludes the presence of Zn in the metal lic state and confirms that the Zn exists only in the oxidized state [41]. For O 1s peak of undoped ZnO and 2% Al:ZnO, the low binding energy component located at 530 eV (Gaussian fit 1) is ascribed to $O^{2-}$ ions attached to $Zn^{2+}$ ions in the wurtzite structure of zinc oxide [42]. The second peak positioned at a high binding energy of 531.9 eV (Gaussian fit 2) corresponds to the presence of specific species like absorber oxygen ($CO_3$, $H_2O$ and $O_2$) and hydroxyl groups [43,44]. For more specification of the components localized at 531.9 eV of B.E, full width at half maximum (FWHM) of the peaks can be used as indicators of chemical states changes and physical influences [45]. The FWHM of the peaks positioned at 531.9 eV was 1.8 eV and 1.49 eV for undoped ZnO and 2% Al:ZnO, respectively. The peak positioned at 531.9 eV with FHWM of 1.8 eV is related to the presence of oxygen in OH species [46]. The peak positioned at 531.9 eV with FHWM of 1.49 eV is related to the presence of oxygen in OH of AlOOH species [47]. The presence of these species on the surface is expected because no etching was applied before the analysis. The peak observed at a B.E of 74 eV for 2% Al:ZnO correspond to Al 2p3/2. The binding energy of Al. in the metallic state and Al 2p3/2 in stoichiometric $Al_2O_3$ are 72.2 eV and 74.6 eV, respectively [48]. No peak was observed at these energies. This result indicates that aluminum exists only on $Al^{3+}$ form and integrates successfully into the zinc oxide matrix [44].

## 3.2. Optical spectroscopy properties

The transmission spectra of undoped and Al doped ZnO nanostructures are shown in Fig. 6. The transmittance decreased with increasing doping concentration. The increase of light absorption related to an increase of free electron concentration in the sample as a result of more incorporation of $Al^{3+}$ ions [49].

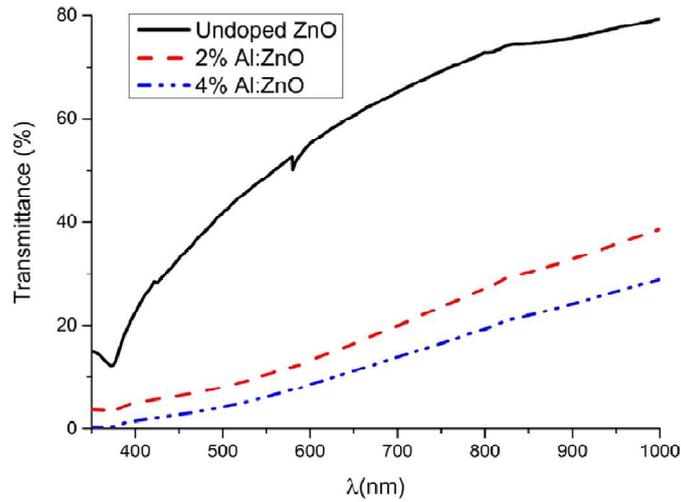

Fig. 6. Optical transmission spectra of undoped and Al doped ZnO nanostructures

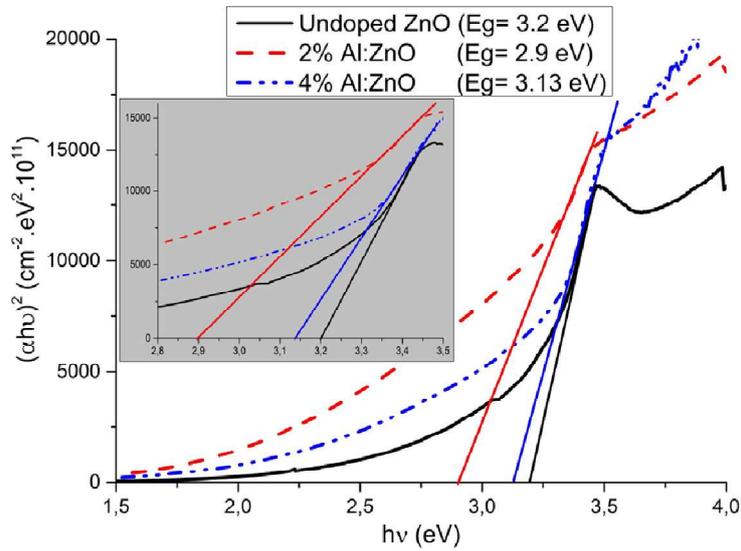

Fig. 7. Optical band gap calculated using the Tauc formula for undoped and Al. doped ZnO nanostructures

The optical band gap was determined using the Tauc formula [50] and transmission data. The procedure for energy band gap calculations has already been reported elsewhere [51,52]. Fig. 7 shows the variation of $(\alpha h\nu)^2$ according to energy ($h\nu$) for undoped and Al doped ZnO nanostructures. The optical band gap values show a decrease from 3.2 eV to 2.9 eV for undoped ZnO and 2% Al:ZnO samples, respectively. Further increase in Al concentration to 4% caused a widening of the optical band gap to 3.13 eV, as illustrated in the inset of Fig. 7. The decrease in the optical band gap is caused by the distortions in the ZnO crystalline matrix due to the incorporation of $Al^{3+}$. This results in an increase in the concentration of free electrons, which leads to the creation of defect levels within the band gap just below the conduction band [53]. The widening of the optical band gap could be described by the start of the Burstein–Moss effect [54], where the Fermi level moves to high energy with an increase in the electron concentration. The low energy transitions are blocked due to the extra electrons introduced by Al doping [55]. A similar result has been reported, in which the optical band gap of ZnO decreased for low Al doping concentration and increased for higher Al doping concentration [56]. The spectral shapes of the SHG and THG are depicted in Fig. 8a and 8b, respectively.

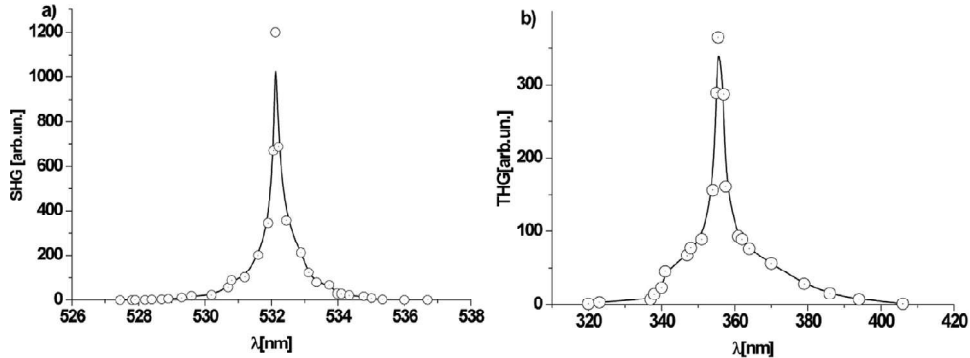

Fig. 8. The spectra of (a) the SHG and (b) the THG after two beams of coherent treatment for 2% Al:ZnO sample

There is observed a drastic jump for the SHG and THG compared to the fluoerescence/ scattering background which confirming that the nonlinear optical origin of the observed effects. The data are obtained after the initial treatment by 60–80 two coherent pulses incident on the 2% Al doped ZnO surface at angles varying within the 25–30°. The earlier studied oxide glasses have been served as reference specimens. These experiments have been carried out for about 60 different points to achieve the necessary statistics and increase the accuracy of the data reproduction.

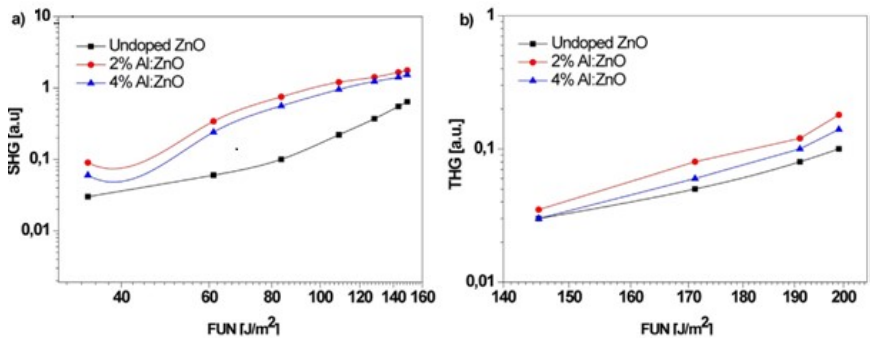

Fig. 9. Dependences of the SHG (a) and (b) THG versus the fundamental laser light energy density

Fig. 9 presents the SHG and THG versus the fundamental laser Light energy density for undoped ZnO, 2% Al:ZnO and 4% Al:ZnO. The presented data confirm that the maximum NLO responses are achieved for the 2% Al:ZnO samples. This could be due to the fact that the distances between nanostructures determined by the concentration of Al. have an optimal value of the ground state and excited dipole moments. The crystalline growth orientation, the shape and packing of nanograins, and optical properties can also have a significant effect. A similar result was found when the NLO properties of Al doped ZnO material reach their maximum for 1% Al content and decrease for high er Al contents (3% and 5%) [28]. It is necessary to emphasize that, for the laser operated THG and SHG, principally different effects were dealt with respect to the traditional NLO effects. In this case, the general phenomenology of the NLO effect will be different and may be described by a phenomenologymodified by external laser induced field (Eqs. (8) and (9)).

$$p_i^{(2)}(\omega) = d_{ijk}^{(2)}(2\omega;\ \omega,\ \omega,\ 0)E_j(\omega)E_k(\omega)E_{ef}(0) \quad (8)$$

$$p_i^{(3)}(\omega) = d_{ijk}^{(3)}(3\omega;\ \omega,\ \omega,\ 0)E_j(\omega)E_k(\omega)E_l(\omega)E_{ef}(0) \quad (9)$$

where dijk, and dijkl are the NLO optical susceptibilities of the second and third order described by third and fourth order polar tensors for SHG and THG, respectively. The principal difference with respect to traditional NLO phenomenological description consisting in the occurrence of the internal dc-electric field Eef(0) occurring due to the interaction of the two coherent beams as described in [37]. Additionally, due to photoinduced changes, some modifications will be appeared in coherent lengths due to laser induced birefringence. The laser stimulated NLO values are at least 10% higher than the doped ZnO nanomaterials synthesized by other methods using the same set-up [14,16,37,57–59] and other borate materials [60,61] measured by the same method.

## 4. Conclusions

The preparation of undoped ZnO and Al doped ZnO nanostructures was successfully achieved using the modified SILAR method. It was found that aluminum doping leads to an improvement in the crystallinity without affecting the basic hexagonal wurtzite structure as indicated in the XRD results. The morphology changes from rod-like in case of undoped ZnO to nanosheet in the presence of aluminum. The EDS and XPS results confirm the presence of Zn, O and Al without the detection of any impurities. As indicated in the compositional analysis, aluminum was successfully incorporated into the ZnO matrix in ionic form (Al3+). This led to an increase of free electron. Therefore, Light absorption was enhanced and the band gap of the nanostructures prepared was reduced. The complex laser stimulated studies of the undoped and Al moped ZnO nanostructures are done. They highlight the crucial role of Al in the properties of ZnO nanostructures. It was found that Al doping caused an enhancement of the SHG/THG which achieves its maximum at 2% Al. content. The laser stimulated NLO values are at least 10% higher than the doped ZnO nanostructures synthesized by other methods using the same set-up. The Al doped ZnO nanostructures prepared can be applied in the field of optoelectronic and photocatalysis due to the high surface area and the increase of the charge density and light absorption.

**CRediT authorship contribution statement**

Ilyass Jellal: Conceptualization, Methodology. Khalid Nouneh: Supervision, Writing - original draft. Jaroslaw Jedryka: Investigation, Formal analysis. Denis Chaumont: Resources, Writing - review & editing. Jamal Naja: Project administration.

**Declaration of Competing Interest**

The authors declare that they have no known competing financial interests or personal relationships that could have appeared to influence the work reported in this paper.

**Acknowledgements**

The authors would like to acknowledge the support through the R& D Initiative – Call for projects around phosphates APPHOS – sponsored by OCP Morocco under the project entitled *Development of a phosphate- based photocatalytic reactor prototype for the treatment and recycling of wastewater*, project ID: TRT-NAJ-01/2017. The authors acknowledge also the European Union's Horizon 2020 research and innovation program under the Marie Skłodowska-Curie grant agrement No 778156 (Project N° W13/H2020/2018). The authors thanks Professor Bernabe Mari Soucase and Hanae Toura from the School of Design Engineering, Universitat Politecnica de Valencia Spain, for some characterizations of the samples prepared.

**Appendix A. Supplementary material**

Supplementary data to this article can be found online at

https:// doi.org/10.1016/j.optlastec.2020.106348.